# Preparing Reproducible Scientific Artifacts using Docker

Michael Canesche, Roland Leißa and Fernando Magno Quintão Pereira*

August 27, 2023

**Abstract**

The pursuit of scientific knowledge strongly depends on the ability to reproduce and validate research results. It is a well-known fact that the scientific community faces challenges related to transparency, reliability, and the reproducibility of empirical published results. Consequently, the design and preparation of reproducible artifacts has a fundamental role in the development of science. Reproducible artifacts comprise comprehensive documentation, data, and code that enable replication and validation of research findings by others. In this work, we discuss a methodology to construct reproducible artifacts based on Docker. Our presentation centers around the preparation of an artifact to be submitted to scientific venues that encourage or require this process. This report's primary audience are scientists working with empirical computer science; however, we believe that the presented methodology can be extended to other technology-oriented empirical disciplines.

## 1 Introduction

The pursuit of scientific knowledge hinges upon the ability to reproduce and validate research findings. Thus, as the scientific community grapples with issues of transparency, reliability, and reproducibility, the importance of preparing research artifacts has come to the forefront [4, 9]. Within the discipline of Computer Science, reproducible artifacts encompass the comprehensive documentation, data, and code that enable the replication and validation of research work by others [11].

In this context, in 2015, the Special Interest Group in Information Retrieval (SIGIR) took the initiative to implement a process that would be, henceforth, known as the ACM Artifact Review and Badging [8, 1]. Artifact evaluation was quickly incorporated into the common practices of another ACM special interest group: SIGPLAN, which focus on research in the field of Programming Languages. Since then, the process of submitting a paper to conferences such as OOPSLA (*Object-oriented Programming, Systems, Languages, and Applications*), PLDI (*Programming Languages, Design and Implementation*), GGO (*Code Generation and Optimization*), CC (*Compiler Construction*) and many others has been done in conjunction with the addition of artifacts. The implementation of ACM Artifact Review and Badging has significantly contributed to the adoption of reproducible artifacts in computing research, a practice that, as noticed by Bajpai et al. [2], brings together many benefits:

- **Transparency and Replicability**: Reproducible artifacts allow other researchers to validate and replicate previous work. By providing detailed information about the methodology, data, and code used, authors enable others to independently verify findings or build upon previous efforts.

- **Peer Review and Collaboration**: Reproducible artifacts facilitate the peer review process. When submitting a paper or research project, including reproducible artifacts allows reviewers to assess the validity of the work more effectively. By providing access to the underlying data and code, reviewers can examine the methodology, detect errors or potential biases, and provide valuable feedback.

- **Error Detection and Debugging:** In the effort of replicating previous work, independent researchers might encounter issues that were overlooked by the primary investigators. Therefore, reproducible artifacts increase the chances of mistakes being discovered and fixed.

*Author info: *Michael Canesche*, UFMG, Brazil: michael.canesche@gmail.com; *Roland Leissa*, Uni-Mannheim, Germany: leissa@uni-mannheim.de; *Fernando Pereira*, UFMG, Brazil: fernando@dcc.ufmg.br.

- **Long-Term Accessibility:** Reproducible artifacts ensure long-term accessibility and preservation of research outputs. Publicly available artifacts can be accessed and utilized by the broader scientific community, even years after their creation. This accessibility promotes the cumulative nature of scientific progress and avoids the loss of valuable knowledge.

- **Education and Training:** Reproducible artifacts are valuable educational resources that contribute to the training of future scientists. They provide a practical and tangible way for students, early-career researchers, or individuals interested in a particular field to learn and understand complex methodologies.

- **Reproducibility Crisis Mitigation:** Reproducibility is an ongoing concern in many scientific fields. By actively preparing artifacts, researchers contribute to mitigating the so-called *Replication Crisis* [3]. Rigorous documentation, open data, and well-organized code address concerns regarding irreproducibility and increase the overall credibility and reliability of scientific research.

In computer science, there are many ways to prepare reproducible artifacts. These artifacts are, typically, collections of scripts, tools, library files and data that replicate tables and charts available in published papers. This document shares our experience using a particular framework, Docker, to prepare and submit artifacts to different ACM artifact evaluation committees. Docker is a system that supports the creation of *containers*—bundles of software that contain all the dependencies necessary to the execution of some tool. This form of virtualization simplifies the process of building reproduce tools and speedups up the process of construction of artifacts.

## 2 Artifact Review and Badging

Several ACM conferences have instituted formal workflows for artifact submission and evaluation. This process is called *Artifact Review and Badging*. Usually, authors of accepted papers are invited to submit artifacts, which will be evaluated by an independent committee. The conference announces the submission platform such as `Hotcrp`. The specific details and requirements of the submission process can vary depending on the venue. Nevertheless, Figure 1 shows the main steps that are often part of the artifact submission process:

1. **Preparation:** An artifact encapsulates a number of scientific experiments. Section 3.1 of this tutorial describes such an experiment. Before submitting an artifact, it's important to ensure that the experiments that it encapsulates meet standard requirements of reproducibility. This may involve performing quality checks, such as code reviews, testing, and documentation, to ensure that the artifact is complete, functional, and well-documented.

2. **Packaging:** The artifact may need to be properly packaged or bundled, depending on the submission requirements. This could involve creating a compressed archive, a distributable package, or a containerized version of the artifact. The packaging process may also involve including any necessary dependencies, licenses, or documentation files. Section 3.2 of this tutorial shows how to package an artifact using Docker.

3. **Submission Platform:** Identify the platform or system where the artifact needs to be submitted. This could be a version control system, a software repository (such as `GitHub`, `GitLab`, or `Zenodo`), or a specific submission portal provided by the organization or project (such as `Hotcrp` or `EasyChair`).

4. **Account Setup:** If the submission requires an account or registration, create the necessary account and provide any required information or credentials. This step ensures that authors have the necessary permissions and access to submit the artifact.

5. **Submission Process:** The submission process is specified by the platform or organization. This typically involves providing information about the artifact, such as its name, version, description, and any associated metadata. Authors may need to upload the artifact file or provide a link to its location. Section 4 provides some guidelines on how to submit an artifact to a SIGPLAN evaluation committee.

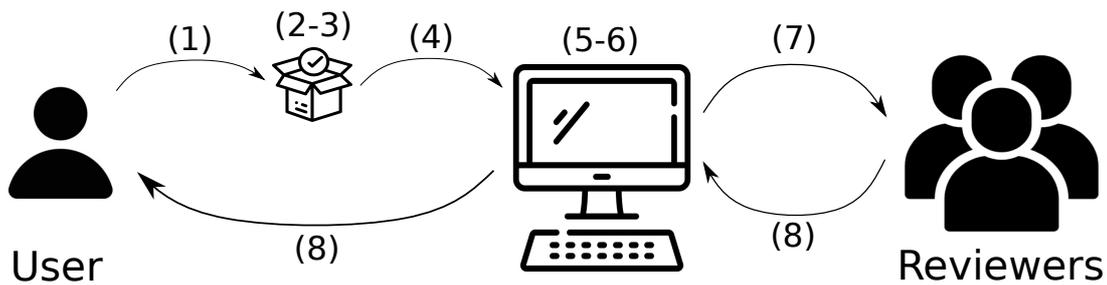

Figure 1: Artifact submission process.

6. **Documentation and Metadata:** Depending on the submission requirements, authors may be asked to provide additional documentation or metadata about the artifact. Metadata could include a `README` file, installation instructions, usage examples, or licensing information. Authors must ensure that all required information is provided accurately and comprehensively. Appendix A of this tutorial contains an example of a typical artifact checklist submitted alongside a paper accepted for publication.

7. **Review:** After submitting the artifact, it goes through a review process, where independent referees try to reproduce the scientific results packaged in the artifact. This process could involve manual execution of the scripts in the artifact, as Section 3.3 of this tutorial explains. Reviewing ensures that the artifact meets necessary quality, security, or compatibility standards. Reviewing is typically *blind*: Authors do not know who the reviewers are. The process can also be *double-blind*: In this case the reviewers also do not know the identity of the authors.

8. **Feedback and Iteration:** In some cases, feedback or requests for changes may be provided during the review process. At this point, authors might have the chance to address any feedback, make the required changes, and resubmit the updated artifact. To ensure anonymity, such interactions can happen through the mediation of the *Artifact Evaluation Chair*, an individual in charge of coordinating the evaluation committee.

If the artifact is approved in this reviewing process, a number of actions shall follow:

**Badging:** The paper that inspired the artifact might receive badges. These are digital stamps that certify different properties of the artifact, such as *availability* and *replicability* of results.

**Publication:** The artifact can be made publicly available in the ACM Digital Library, where interested researchers can download and use it. In this case, the artifact receives a Digital Object Identifier (DOI)—a number that uniquely identifies the entry in the digital library.

## 3 Tutorial: Writing an Artifact with Docker

This section presents a best-practices tutorial on how to prepare an artifact using Docker.[1] The section follows the steps outlined in Figure 2. The figure emphasizes that artifacts are *products* of empirical research. As such, they exist to reproduce *experiments*. Section 3.1 describes one such experiment. The author of the experiment—the *primary researcher*—prepares scripts to generate the *results* of these experiments: charts and tables that summarize the observations that the experiment enables. The process of construction of an artifact is the subject of Section 3.2. A Docker artifact requires a *Dockerfile*, which Section 3.2.1 describes. Once an artifact is ready for use, it can be downloaded by a *secondary researcher*: someone interested in reproducing the scientific results earlier produced by the primary researcher. Section 3.3 describes the process to reproduce the experiments packaged in the Docker artifact.

---

[1]Throughout this tutorial, we shall use `GitHub` to make the artifact available (Link: https://github.com/lac-dcc/koroghlu). Whoever follows this tutorial must have `docker`, `docker-compose`, and `git` installed.

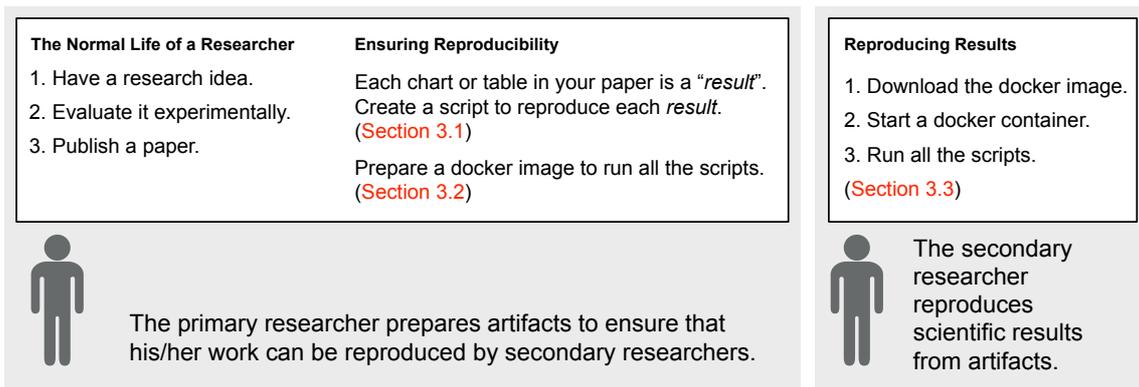

Figure 2: The process of packaging empirical results into reusable artifacts.

**The Docker Glossary.** Docker is a platform that enables developers to create, deploy, and run applications in a *containerized* environment. *Containers* are lightweight, portable, and self-contained software packages that include everything an application needs to run, such as libraries, dependencies, and configuration files. In general, Docker simplifies the process of creating and deploying an application, because the same software resources in which the primary researcher developed the tool will be used by the secondary researcher to replicate the result. Docker provides a way to package and distribute applications as containers, allowing them to run consistently across different environments, from development to production [10]. Docker works by using a *layered file system* to build and store containers. Each layer represents a specific piece of an application's environment, such as the operating system, libraries, or application code. These layers can be combined and reused to create new containers. Docker also provides a set of tools and APIs for managing containers, including the *Docker Command Line Interface* (CLI), *Docker Compose*, and *Docker Swarm*. These tools allow developers to build, test, and deploy applications using containers, while also providing features for scaling, load balancing, and service discovery. This tutorial covers the usage of the command line interface.

## 3.1 The Experiment

An artifact reproduces the results observed in a scientific experiment. Thus, this tutorial requires an experiment, which is a set of procedures that answer some research question. As an example, we shall investigate the following research question:

> **RQ:** What is the relative search time of the different auto-tuning algorithms available in Apache AutoTVM [5]?

The nature of our research question is immaterial to the understanding of this tutorial. Nevertheless, we provide some explanation on the terms that it mentions. Apache AutoTVM is a module within Apache TVM, a specialized compiler, written as a set of Python libraries, that optimizes machine-learning models. A machine-learning model, such as ResNet, MobileNet or RXNet can be understood as a chain of *kernels*: algorithms that apply some computation on an ensemble of data. There are many different ways to generate binary code for a given machine-learning model. The problem of finding the best in such a way is called *kernel scheduling* or *program autotuning*. AutoTVM has different algorithms to find these best implementations. In this paper we analyze four of them:

**GridSearch:** Explores exhaustively a bounded number of model configurations. Regular ranges of transformation parameters determine this set of configurations.

**Random:** Samples different models randomly. Sampling usually follows a uniform distribution on predefined bounds placed onto the parameters.

**GA:** Uses a genetic algorithm to steer the search for good model configurations. The parameters of previous models are used to find the parameters of the next best model candidates.

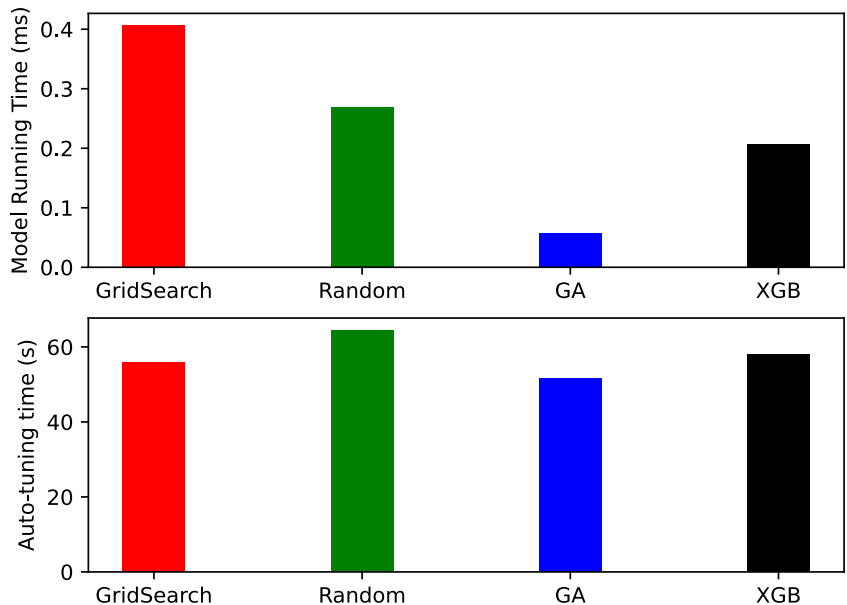

Figure 3: (Top) The running time of the models produced by different auto-tuning techniques available in AutoTVM. (Bottom) The time that each auto-tuning approach takes to explore the space of model implementations.

**XGB:** Uses simulated annealing as a refinement of random search, where sampling alternates between regions that are close and distant from current best points.

These different algorithms provide a tradeoff between search time and the efficiency of the final model that they produce. Figure 3 compares the relative merits of each approach. This figure is our *scientific result*, as mentioned in Figure 2. This tutorial explains how to set up an artifact to reproduce this result. The experimental setup used to produce the original version of the figure is given below:

**Hardware:** AMD Ryzen 7 4800H processor with 8 kernels, and 32 GB Ram memory

**Software:** Linux Ubuntu 20.04; Apache TVM 0.10 and Python 3.8.2

**Benchmark:** matrix multiplication using four different search models[2]

Throughout the rest of this tutorial, we shall assume that the artifact is organized according to the structure described in Figure 4. Thus, our artifact is formed by a suite of scripts (the light gray boxes in Figure 4) plus a number of Dockerfiles (the dark gray boxes). We shall call the original scripts the *primary* artifact products: these are the items that were developed to answer the research question. The Dockerfiles are *meta* artifact products: they do not exist to generate the scientific results, but rather to reproduce them from the primary items.

## 3.2 Building an Artifact with the Docker Command Line Interface

The *Docker Command Line Interface* (CLI) is a suite of commands that let users build containers through the prompt of an operating system's terminal. The following steps build a Docker image with everything necessary to reproduce the experiment in Section 3.1 using the command line interface of Docker:

---

[2]Link to the benchmark: `https://github.com/lac-dcc/koroghlu/blob/main/src/mm.py`

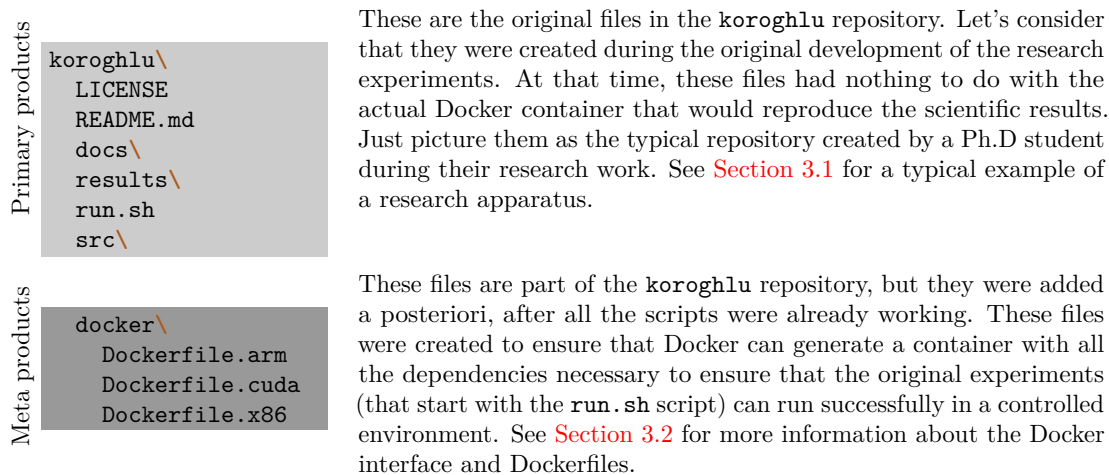

| Primary products | `koroghlu\`<br>`  LICENSE`<br>`  README.md`<br>`  docs\`<br>`  results\`<br>`  run.sh`<br>`  src\` | These are the original files in the `koroghlu` repository. Let's consider that they were created during the original development of the research experiments. At that time, these files had nothing to do with the actual Docker container that would reproduce the scientific results. Just picture them as the typical repository created by a Ph.D student during their research work. See Section 3.1 for a typical example of a research apparatus. |

| Meta products | `docker\`<br>`  Dockerfile.arm`<br>`  Dockerfile.cuda`<br>`  Dockerfile.x86` | These files are part of the `koroghlu` repository, but they were added a posteriori, after all the scripts were already working. These files were created to ensure that Docker can generate a container with all the dependencies necessary to ensure that the original experiments (that start with the `run.sh` script) can run successfully in a controlled environment. See Section 3.2 for more information about the Docker interface and Dockerfiles. |

Figure 4: The structure of the repository that contains the artifact described in this tutorial.

1. Install the `docker` tool following the official documentation[3]. As an example, on Linux or in the Windows WSL the following command should be enough:

   ```
   $ sudo apt install docker.io
   ```

   On OSX, the following command could be used instead:

   ```
   $ brew install docker
   ```

2. Download the code necessary to run the experiment described in Section 3.1:

   ```
   $ git clone https://github.com/lac-dcc/koroghlu
   ```

3. Move onto the `koroghlu` folder, which contains the build scripts:

   ```
   $ cd koroghlu/
   ```

4. Build a Docker image by running the command below within the `koroghlu` folder. This command builds an image from a Dockerfile, whose access path is specified with the `-f` tag (Estimated build time: **10 minutes** on an Intel machine with 2.8GHz of clock):

   ```
   $ docker build -t docker-artifact -f docker/Dockerfile.x86 .
   ```

   The previous command builds a Docker image with the "tag" `docker-artifact` (specified after the `-t` flag). The tag is a name (of our own choice) that we shall use to refer to this image in other commands.

   **Remark.** The Docker daemon accesses a Unix socket owned by the `root` user. Thus, depending on privileges, users might have to run Docker commands as `sudo`. To avoid prefixing Docker commands with `sudo`, create a Unix group called `docker`. Users in this group will be able to run `docker` without root access. To follow this path, do:

   ```
   $ sudo groupadd docker
   $ sudo usermod -aG docker $USER
   ```

At the end of this forth step, a Docker image is created. This image follows the specifications given in `Dockerfile.x86`. This file is the core of the artifact, as Section 3.2.1 shall explain.

---

[3]Available at https://docs.docker.com/engine/install/ on June 9th, 2023

### 3.2.1 The Dockerfile

To prepare the artifact evaluated in Section 3.2, two items are necessary:

1. A git repository containing all the scripts that reproduce the experiment. The repository used in this tutorial is available at `https://github.com/lac-dcc/koroghlu`.

2. A Dockerfile. We provide three different files in the folder `koroghlu/docker`, which is part of the git repository. Each file builds a Docker image to reproduce experiments using the setup of a specific computer architecture (`arm`, `cuda` or `x86`). The rest of this section explains how the Dockerfile of `x86` is organized.

A reproducible artifact evaluated through Docker consists of a *Docker Image*: a file used to execute code in a container. The *Dockerfile* is a text file containing the commands used to build a Docker image. The Dockerfile typically starts with a base image, which serves as the foundation for the container. It also includes a series of instructions that specify how to install dependencies, configure the environment, copy files into the container, and reproduce experiments. Listing 1 shows how a typical Dockerfile is organized. Notice that a Dockerfile is structured around *instructions* such as `RUN`, `WORKDIR`, etc. explained below.

Listing 1 shows the Dockerfile that this tutorial provides for the x86-64 architecture[4]. In this example, the image is provided by the Ubuntu 20.04 server, as seen in Line 2 of Listing 1. The commands at Lines 5 and 13 configure the date, time and language of the system. The `RUN` instructions install system dependencies (Lines 9-10, and 16-18) and program dependencies (Line 20). They also install and process data available in public repositories (Lines 27-33). The `WORKDIR` statement at Line 36 sets the folder where all these commands will run. Finally, the `ENTRYPOINT` instruction, at Line 37, defines the primary executable for this artifact.

Listing 1 uses Dockerfile instructions—special keywords—to describe how the artifact can be reconstructed and executed. For the sake of completeness, we summarize the keywords used in Listing 1 below. However, notice that this tutorial covers a small number of Dockerfile keywords. For a complete list, we refer the reader to the official Docker documentation[5]:

`FROM`: This instruction specifies how to download the image from a publicly available server.

`ENV`: This instruction declares environment variables and defines their values. These variables can be read by any process running inside the container, such as the application or services that the container encapsulates.

`RUN`: This instruction executes commands during the build process of a Docker image.

`WORKDIR`: This instruction sets the working directory for any subsequent instructions in the Dockerfile. Thus, it defines the directory where commands loaded via `RUN` instructions (or similar tags, such as `CMD`, `COPY`, and `ADD`) will execute.

`ENTRYPOINT`: This instruction specifies the command that will be executed when a Docker container starts. Thus, it defines the primary executable for the container as seen in Line 37.

## 3.3 Reproducing a Scientific Result within a Docker Image

Scientific results can be reproduced from a Docker image. As explained in the beginning of this section, from the Docker image, containers can be created. Indeed, from one image, multiple containers can be instantiated. As an analogy borrowed from object-oriented programming, a Docker image relates to a "class" as a Docker container relates to "objects". Docker images can be stored for download, for instance. In fact, Docker gave origin to *Docker Hub*[6], a publicly available repository of images, which can be freely downloaded and reused. In this tutorial, we assume that you have a ready-to-use Docker image. Such an image will be generated at the end of step 4 in Section 3.2. To reproduce the experiment packaged in that image, proceed as follows:

---

[4] Notice that our tutorial contains additional files for ARM (`Dockerfile.arm`) and Cuda (`Dockerfile.cuda`).
[5] Available at `https://docs.docker.com/engine/reference/builder/`.
[6] Accessible at `https://hub.docker.com/`.

```dockerfile
######################################## Image ########################################
FROM ubuntu:20.04

##################################### Date and Time ####################################
ENV TZ="America/Sao_Paulo"
RUN ln -snf /usr/share/zoneinfo/$TZ /etc/localtime
RUN echo $TZ > /etc/timezone && rm -rf /var/lib/apt/lists/*

RUN apt-get update -y
RUN apt-get install -y locales curl wget tar sudo git apt-utils
RUN localedef -i en_US -c -f UTF-8 -A /usr/share/locale/locale.alias en_US.UTF-8

ENV LANG en_US.utf8

##################################### DEPENDENCIES #####################################
RUN apt-get install -y gcc g++ graphviz vim python3 python3-pip python3-dev automake \
    make clang build-essential cmake llvm-dev cython3 python-is-python3 libedit-dev \
    libtinfo-dev python3-setuptools libxml2-dev

RUN pip3 install numpy==1.20 decorator scipy pytest psutil typing_extensions synr \
    tornado cloudpickle 'xgboost<1.6.0' mxnet pandas matplotlib

##################################### COPY ARTIFACT ####################################
RUN cd $HOME && git clone https://github.com/canesche/docker-artifact

##################################### INSTALL TVM #####################################
RUN cd $HOME && git clone -b v0.10.0 --recursive https://github.com/apache/tvm && \
    cd tvm && mkdir -p build && \
    cp $HOME/docker-artifact/docker/config.cmake.x86 $HOME/tvm/build && \
    mv $HOME/tvm/build/config.cmake.x86 $HOME/tvm/build/config.cmake && \
    cd build && cmake .. && make -j8 && cd .. && sudo make cython3

RUN echo "export PYTHONPATH=/root/tvm/python:/python:" >> ~/.bashrc

##################################### SET WORKDIR #####################################
WORKDIR /root/docker-artifact
ENTRYPOINT ["/bin/bash"]
```

Listing 1: Example of a Docker file that reproduces experiments on the x86 setup.

1. Once a Docker image is built with the tag name docker-artifact, run the artifact that this image contains with the following command:

    $ docker run -ti -v ${PWD}/results:/root/koroghlu/results docker-artifact

2. In the Docker prompt, execute the command below to reproduce the experiments (Estimated running time: **5 minutes** on an Intel machine with 2.8GHz of clock.):

    root@f1258685f4fd:~/koroghlu# ./run.sh x86

    **Remark.** The run.sh script will run AutoTVM within the Docker container. The script prints out which search approach it is currently using (GridSearchTuner, RandomTuner, etc). If AutoTVM is unable to find any valid schedule for a given task, then it will print out a warning message as well. These warnings will not prevent AutoTVM from finding valid schedulers.

3. Once the `run.sh` script terminates, we must have results ready to be analyzed in the `results` folder. We can exit the Docker container and enter that folder to check out the results that we have reproduced:

```
$ root@f1258685f4fd:~/koroghlu# exit
$ cd results/
$ cat results.csv
  Turning,time(ms),std(ms),Space search(s),tile_i,tile_j,tile_k,order
  GridSearchTuner,1.0928,0.0589,103.33
  RandomTuner,0.6074,0.0000,95.20,64,1,16,jki
  GATuner,0.1129,0.0004,69.99,1,80,1,jki
  XGBTuner,0.3067,0.0002,84.30,16,48,64,jki
$ ls *.pdf
  x86_tuning.pdf
```

At the end of the third step above, we must have a PDF figure in the `results` folder: `x86_tuning.pdf`. This figure should be similar to Figure 3: it represents the "scientific result" produced by our artifact. Notice that the same artifact can be used to produce several different scientific results. A good practice is to have a separate script (like our `run.sh` above) to reproduce each one of these results.

## 4 The Artifact Evaluation Process

As mentioned in Section 1, artifact evaluation is common practice, not only in ACM SIGPLAN conferences, but also in venues outside the umbrella of the Association for Computing Machinery, such as the USENIX Organization or the NeurIPS Foundation. Most of the conferences (and some journals) that offer artifact evaluation follow the same two steps:

**Paper submission:** Papers are submitted and evaluated by a *Program Evaluation Committee*.

**Artifact submission:** Authors of papers accepted in the first phase are invited to submit artifacts, which will be evaluated by an *Artifact Evaluation Committee*.

Usually, the successful evaluation of a paper's companion artifact is not a requirement for the paper to be published in the conference's proceeding. Acceptance by the program evaluation committee is considered a sufficient condition for publication. Therefore, authors might or might not submit artifacts for evaluation. Nevertheless, the submission of an artifact brings all the advantages already enumerated in Section 1 of this tutorial: transparency, replicability, error detection, accessibility, etc. However, some conferences call for special *Tool* or *Practical Experience Papers*. With these, it is sometimes necessary to achieve a minimum level of confidence in the artifact evaluation process. This section of the tutorial covers the submission and evaluation process.

### 4.1 The Artifact Evaluation Committee

For the main conference, the *Program Chair* (PC Chair) organizes and recruits international, expert scientists for the *program committee*, who are responsible for reviewing papers. Likewise, the *Artifact Evaluation Chair* (AE Chair) organizes and recruits international researchers for the *artifact evaluation committee*, who are responsible for reviewing artifacts. While the program committee usually consists of professors and senior developers, the artifact evaluation committee typically consists of Ph.D students and young researchers. The reason for this is that reproducing artifacts itself requires less scientific and more technical knowledge. It is not uncommon that people with the appropriate skills and background knowledge ask the chair to participate in the evaluation committee (the AE chair often gratefully accept any help). This is a great opportunity for Ph.D students and anyone interested to get in touch with the research community and to learn "the other side" of the reviewing process. Participation is also a good addition on the CV. Most venues keep the artifact evaluation committee public on some web page. However, artifact authors do usually not know who will review their works. The anonymous reviewing process shall be discussed in Section 4.2.

## 4.2 Double-Blind Submission

Many journals and conferences nowadays adopt a "double-blind" submission policy. The double blind submission process ensures anonymity during the review process. In a double blind review, the identities of both the authors and the reviewers are concealed from each other. This format helps eliminate biases that may arise from knowing the identity of the authors, such as their reputation or affiliation. Typically, artifact submission follows this same process of double-blind submission, which consists of the following steps:

1. Authors of accepted papers prepare their artifacts—for instance by following the guidelines in Section 3.2. Meta information about the artifact is appended to the anonymous paper. This metadata should not include information about the authors. For an example of metadata, see Appendix A at the end of this tutorial.

2. The anonymized paper, with the artifact metadata, is submitted to the conference management system. The system assigns a unique identification number to the paper for tracking purposes.

3. The program committee chair assigns reviewers to evaluate the paper's artifact. There referees are drawn from the program committee, usually based on familiarity with the research topic. The reviewers are not aware of the authors' identities. Similarly, the authors do not know who will be reviewing their work.

4. Reviewers evaluate the submitted artifact, for instance, following the steps in Section 3.3 of this tutorial. The evaluation is solely based on the content and quality of the research. Reviewers are expected to write a review, where they provide feedback, critiques, and recommendations for improvements, as Section 4.3 discusses.

5. The program committee chair evaluates the reviews and makes decisions regarding acceptance, rejection, or revisions. If necessary, the chair might arrange some form of communication between authors and referees to preserve the anonymity of both parties. These interactions might be necessary, for instance, so that authors can fix small mistakes made during the preparation of the artifact.

6. Once the review process is complete, the authors are notified about the outcome of their submission, usually without disclosing the identities of the reviewers. At this point, the paper might receive one or more badges, as explained in Section 4.4.

## 4.3 The Evaluation Process

Like the process of submitting artifacts (see Section 2), the process of evaluating artifacts happens in multiple steps. In general lines, the following phases are likely to be present in most settings:

1. **Recruiting:** The Artifact Evaluation Chair (AE Chair) assembles the Artifact Evaluation Committee (AEC). Recruiting is done in many ways. For instance, the PC chair may invite people directly, or announce the committee in forums, and ask the interested researchers and practitioners to get in touch.

2. **Profiling:** Once the AEC is formed, the AE chair invites members to fill up their individual profile. This step gives the PC Chair information to help him/her assign papers to referees needs and expertises.

3. **Bidding:** Once artifacts are submitted and stored in the conference system, reviewers have the opportunity to look into all submissions as well as their associated papers, in order to judge which artifacts they feel confident to review.

4. **Assignment:** The AE Chair assigns artifacts to reviewers. Typically, a referee will not obtain more than three reviews. The conference system automatically assigns reviewers to artifacts while considering profile information and bidding scores. On top of that, the AE Chair typically double-checks these assignments and manually tweaks the result.

5. **Evaluation:** The reviewer downloads the artifact and reads the provided documentation as well as the associated paper. These instructions should be enough to roughly reproduce the numbers presented in the paper. Some artifacts require confidential material. In order to not exclude these artifacts, reviewers gain remote access (e.g. via SSH) to a machine where everything is already pre-installed and set up. At this point, the reviewer might have the chance to interact with the artifact's authors. Interactions typically happen anonymously (see Section 4.2).

6. **Writing:** Once evaluation is over, the reviewer writes a brief assignment of his/her impressions on the process. This feedback will help authors to improve the artifact, possibly making it more functional and reusable. Many conference systems also use a grading system. At this stage, the reviewer can set a score to the artifact.

7. **Rebuttal:** At the end of the evaluation phase, the AE Chair notifies the authors, via the conference system, of the outcome of the evaluation. Authors can read the—usually multiple—reviews concerning their artifact, and have the chance to answer questions and clarify misunderstandings in a rebuttal letter. Some venues foster a more fluent discussion between reviewers and authors where the reviewers can directly clarify any problems with the artifacts with the authors.

8. **Discussion:** After rebuttals are in place, reviewers can discuss among themselves the final outcome of the evaluation process. Discussions can be anonymous or not, depending on the conference system, and on previous agreements between the AE Chair and the Conference Steering Committee. If discussions are anonymous, then they happen through a chat system. Non-anonymous discussions can happen either asynchronous, e.g., through a chat system, or through a synchronous meeting (usually online).

9. **Badging:** The AE Chair uses the outcome of discussions and scores to assign badges to artifacts. Different conferences use different badging systems. Badges indicate, for instance, that the artifact was successfully evaluated, or that the artifact is publicly available.

10. **Notification:** After reviewers reach a consensus about each artifact, the AE Chair notifies the authors about the outcome of the evaluation. At this point, authors receive instructions on how to submit the final version of the artifact and how badges that acknowledge that work will be put on the paper.

11. **Archival:** Authors submit a final version of the artifact. For some venues, the authors add the earned badges themselves on the final version of their paper. In other venues, the editors of the journal/proceedings later affix the badges to the camara-ready version of the paper.

Although every step above is important, the bidding stage deserves special attention. Many artifacts have special hardware or software requirements such as access to a special GPU or specific OS. As the bidding process only lasts a few days—sometimes even just 24 hours—reviewers will not have the time to look at all the submissions in detail. For this reason it is important for the authors of a submission to describe all necessary background knowledge as well as hardware, software, time, or space requirements in the abstract of the submission. For example, if the artifact uses the JVM (Java Virtual Machine) to run an experiment that runs several days and produces $\sim 100\,\mathrm{GB}$ of data, authors should clearly communicate this in the abstract so potential reviewers know they should be familiar with the JVM and have access to a machine with enough disk space they can spare for several days.

## 4.4 The Products of a Successful Artifact Evaluation

The result of the artifact evaluation can be made public in two—non-exclusive—ways: *badging* and *archival*.

### 4.4.1 Badging

Badging is a protocol that venues use to indicate to what extent the ideas described in an artifact could be successfully evaluated. Different venues might adopt different badges. As an example, USENIX and ACM adopt at least these three badges:

**Available:** The artifact is publicly available in some archival system (see Section 4.4.2).

**Functional:** The artifact has been judged to be easy to use, practical and well documented.

**Reproduced:** The artifact has been judged to support its authors' claims, being independently executed.

It is common that badges have a graphic representation. This is stamped on the papers and/or added to digital libraries to indicate how the papers' accompanying artifacts have been evaluated. Figure 5 shows examples of badges used by two different venues: the *USENIX Security Conference*, and the *Association for Computing Machinery* (ACM). Note that different venues may use different badges. In particular, ACM assigns two extra badges to papers, not listed in Figure 5: "*Artifact Replicated*" and "*Artifact Reusable*". These indicate, respectively, that the work has been replicated for independent groups after the artifact evaluation process, and that the artifact associated with the paper is of a quality that significantly exceeds minimal functionality.

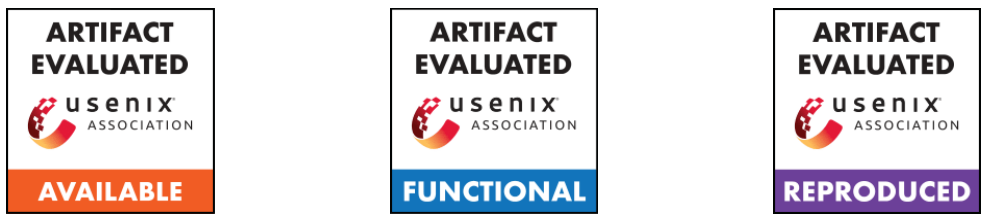

Source [seen in 2023]: `https://secartifacts.github.io/usenixsec2023/badges`

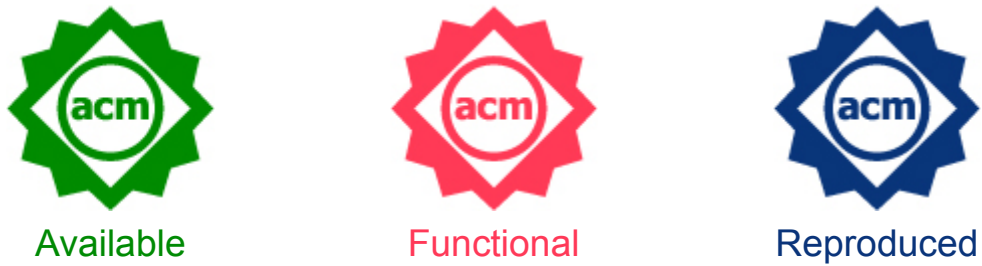

Source [seen in 2023]: `https://www.acm.org/publications/policies/artifact-review-badging`

Figure 5: (Top) Badges used by USENIX Security. (Bottom) Equivalent badges used by the Association for Computing Machinery.

### 4.4.2 Archival

Archival, as hinted in Section 4.3, is the process of making artifacts available for retrieval. This step is not mandatory in every venue that performs artifact evaluation: some software simply cannot be made public, due to commercial concerns or risk of causing harm, for instance. Still, such artifacts can be evaluated, and the results of the evaluation can be reported publicly. On the other hand, if the artifact can be made permanently public, then authors have different alternatives at hand to do it. Examples of such resorts include:

**Digital Libraries:** websites that assign artifacts unique identifiers, and ensure that specific versions of these artifacts are available. Examples include `Zenodo` or the *ACM Digital Library*

**Repositories:** software development repositories like `GitHub` or `GitLab` can be used to store artifacts. These repositories let artifacts evolve over time. In this case, venues might require stable reference identifiers for artifacts.

**Institution:** some universities, companies and research centers maintain websites where artifacts can be stored. Again, venues might require that these systems provide identifiers for stable references of artifacts.

Figure 6 shows an example of paper whose accompanying artifact has been successfully evaluated. This example paper [6] was published in the *International Symposium on Code Generation and Optimization* (CGO'23), a conference that provides authors with artifact evaluation. Upon successful evaluation, the paper receives different badges, which are stamped on its front page, as Figure 6 (Left) shows. The paper's entry on the ACM Digital Library also features the same badges, plus a link to the actual artifact. In this case, the artifact is stored in `Zenodo`, a general-purpose open repository operated by *The European Organization for Nuclear Research* (CERN). `Zenodo`'s items are associated with a persistent *Digital Object Identifier* (DOI). Hence, these artifacts can be cited independently from the papers where they were published. Such is, for instance, the case of the example in Figure 6 [7].

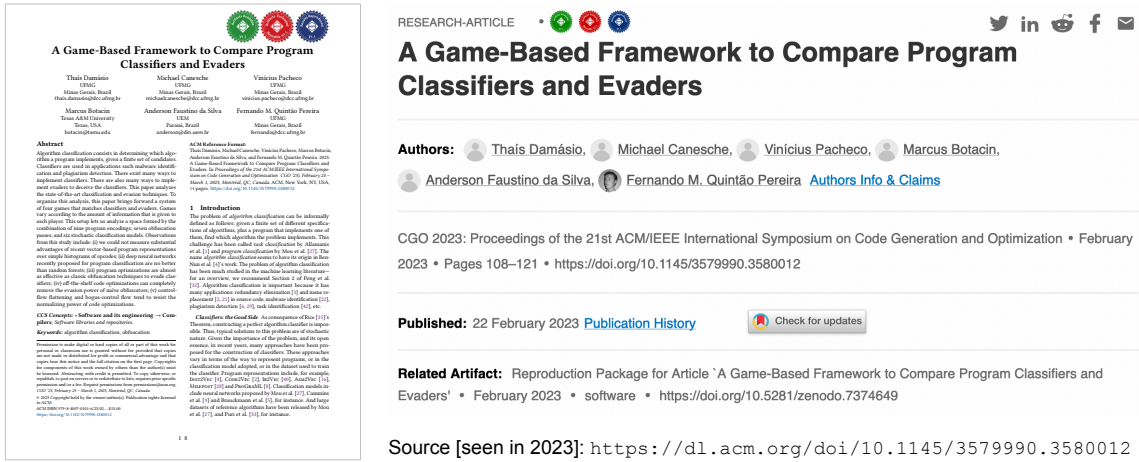

Figure 6: (Left) Example of paper, with artifact evaluation badges on the front page. (Right) Paper's entry on the ACM Digital Library (`https://dl.acm.org/doi/10.1145/3579990.3580012`), with link to artifact stored in Zenodo (at `https://doi.org/10.5281/zenodo.7374649`).

# 5  Conclusion

The integration of artifact submission processes in conferences is enhancing the quality, credibility, and impact of research work. As the scientific community continues to embrace reproducible artifacts, different tools and protocols are being developed to support transparency, collaboration, and robustness in research. This tutorial has covered one such tool: Docker. Overall, using Docker to prepare reproducible artifacts offers significant benefits for artifact creators and evaluators, in terms of consistency, collaboration, and portability. The tutorial is far from being exhaustive. Docker offers several different extensions, such as Docker Compose or Docker Swarm, which we have not touched in this document. `Compose` lets users run applications involving multiple containers; `Swarm` lets users manage clusters of Docker daemons. Additionally, there are several commands available to Docker CLI and many keywords for the Dockerfile language that we have not discussed. To know more about these extensions, we invite the interested reader to check out Docker's official documentation in the tool's webpage.

# A  Artifact Appendix

This appendix simulates the extension usually added to a paper that provides a research artifact. This extension can be used as a checklist that helps the secondary researchers to reproduce the results obtained by the primary researcher.

## A.1 Abstract

This artifact compares different search models available in Apache AutoTVM. In total, this artifact let us evaluate four search models when tuning a standard implementation of the matrix multiplication kernel. The artifact consists of a Docker container with accompanying scripts to replicate Figure 3 automatically.

## A.2 Artifact check-list (meta-information)

- **Benchmark:** Matrix multiplication.
- **Goal:** Reproduce Figure 3.
- **Compilation:** clang, cmake
- **Dataset:** Implementation of matrix multiplication using libraries from Apache TVM.
- **Runtime environment:** Any operating system that supports Docker, Python3, Wget, Tar, and Sed.
- **Hardware:** Any x86-64 machine.
- **Metrics:** Time.
- **Output:** Figure 3 in PDF format, plus the CSV file used to build the figure.
- **Disk space required (approx.):** 10 GB.
- **Time to prepare workflow (approximately):** 30 minutes.
- **Time to complete experiments (approximately):** 5 minutes
- **Publicly available?** Yes
- **Code licenses (if publicly available):** GPL-3.0.
- **Archived (provide DOI)?:** https://doi.org/10.5281/zenodo.8096830

## A.3 Description

### A.3.1 Delivery

https://doi.org/10.5281/zenodo.8096830

### A.3.2 Hardware dependencies

x86-64 processor.

### A.3.3 Software dependencies

Docker, Python3, Jupyter Notebook, Wget, Tar, and Sed.

### A.3.4 Data sets

- `mm.py`: matrix multiplication in TensorIR, the tensor program abstraction in Apache TVM.

## A.4 Installation

1. Download the artifact from https://github.com/lac-dcc/koroghlu.
2. Install Docker and follow the steps in the Section 3.2.

## A.5 Experimental workflow

To execute the experiments, run the script `run.sh` as follows:

```
$ ./run.sh
```

## A.6 Evaluation and expected result

Once `run.sh` terminates, results will be available in two files:

- `koroghlu/results/x86_tuning.pdf`;
- `koroghlu/results/result.csv`.

The PDF is a reproduction of Figure 3 with the data in the CSV file.